# The Kinematics of the Ionized and Molecular Hydrogen in the Starburst Galaxy NGC 253[1]

F. Prada[1,2,5]

A. Manchado[2]

Blaise Canzian[3]

R.F.Peletier[2,4]

C. D. McKeith[1]

and

J. E. Beckman[2]

[1]*Department of Pure and Applied Physics, Queen's University of Belfast, Belfast BT7 1NN, Northern Ireland*

[2]*Instituto de Astrofísica de Canarias,E-38200 La Laguna, Tenerife, Spain*

[3]*U.S. Naval Observatory, P.O. Box 1149, Flagstaff AZ 86002, USA, and Universities Space Research Association*

[4]*Kapteyn Laboratorium, P.O. Box 800, 9700 AV Groningen, The Netherlands*

[5] *Present Address: Max-Planck-Institut für Extraterrestriche Physick, D-85748 Garching, Munich, Germany*

[1]Based on observations at ESO, La Silla, Chile.

## ABSTRACT

Near-infrared $H_2$ 1–0 S(1) and Br $\gamma$ velocity curves along the major axis of NGC 253 have revealed a central velocity gradient that is seven times steeper than that shown by the optical velocity curve. This is interpreted as an optical depth effect due to dust. Approximately 35 mag of visual extinction in the center is required to match the SW side of the optical velocity curve. The spatial variation of the ratio of these lines to the $^{12}$CO ($J = 1 \rightarrow 0$) line is compared among starburst galaxies NGC 253, M82, and NGC 4945 to investigate the excitation mechanism responsible for the $H_2$ 1–0 S(1) line.

## 1. INTRODUCTION

The recent development of near-infrared arrays has made it possible to observe the kinematics of spiral galaxies using near-infrared emission lines. This is important for measuring kinematics of the nuclear regions in galaxies with heavy dust obscuration since dust, which is concentrated in the galactic plane, extinguished the visible light emitted near the center. At visible wavelengths, only the line of sight velocity of the outer regions of the galaxy can be seen. There are as yet few papers in the literature with near-IR kinematic studies of galaxies. McKeith et al. (1993) established the wavelength dependence of observed kinematics for the starburst galaxy M82, where they found a much steeper velocity gradient for the interstellar [S III] $\lambda$ 9069 Å and stellar Ca II IR triplet than for the visible and near-UV lines. This effect is attributable to dust obscuration, as observed also in the dusty inclined galaxy NGC 2146 (Prada et al. 1994).

In this paper, we report near-IR observations of the nearby starburst spiral galaxy NGC 253. The galaxy has already been observed along the major axis in optical emission lines (Ulrich 1978; Muñoz-Tuñon et al. 1993), and in the $J = 1 \rightarrow 0$ transition of molecular CO (Canzian et al. 1988). There is a significant difference between these two data sets. To confirm that this difference is caused by the presence of dust, we have observed NGC 253 in the near-IR, where extinction is approximately one order lower than in the visible spectral region (Rieke & Lebofsky 1985).

## 2. OBSERVATIONS AND DATA REDUCTION

Long slit spectra of NGC 253 were obtained with the IRSPEC spectrograph (Gredel & Moorwood 1991) at the ESO New Technology Telescope at La Silla in March 1994. A $58 \times 68$ SBRC InSb array was used as detector with a pixel size of $2\farcs2$ . The slit was $120''$ long and $4\farcs4$ wide. Seeing was about $1''$; convolved with the pixel size this gives a FWHM of $2\farcs8$ for a point source. A distance of 3.4 Mpc (Sandage & Tammann 1975) for NGC 253 yields a scale of 16.5 pc arcsec$^{-1}$. Two slit positions, one on the major axis (p.a. 52°) and one on the minor axis (p.a. $-38°$) were observed separately at Br $\gamma$ $\lambda 2.1661\,\mu$m and $H_2$ 1–0 S(1) $\lambda 2.1218\,\mu$m. The spectral dispersion was 65 km s$^{-1}$ pix$^{-1}$ with a velocity resolution of 140 km s$^{-1}$. Individual observations of one minute were interleaved with sky frames in the follow sequence: object-sky-sky-object. In total, 14 minutes integrations were made on the major axis and 8 minutes on the minor, in each spectral range. Figures 1 and 2 show the Br $\gamma$ and $H_2$ intensity profiles at different positions along the major axis. The intensity profiles of both species along the minor axis are also plotted in Figures 3 and 4. In table 1,2,3,4 are listed the Br$\gamma$ and $H_2$ line fluxes and LSR radial velocities at different positions taken from the major and minor axis spectra of NCG 253.

Flatfield and spatial distortion correction, sky subtraction, and wavelength and flux calibration were performed using the IRSPEC package within MIDAS, developed by ESO. Flux calibration was achieved by interpolating photometric data from the standard stars HR5056 and HR2294, in the absence of spectrophotometric data.

The radial velocities were measured by fit-ting Gaussian profiles using the IRAF task FITLINES developed by J. Acosta-Pulido at the IAC. Emission line fluxes were determined by adding pixels. Velocity measurement errors combine photon noise, the uncertainty in the continuum level and the overall goodness of fit. The velocity measurements were corrected to the local standard of rest (LSR).

## 3. RESULTS

### 3.1. Near-IR circumnuclear kinematics

We derived Br $\gamma$ $\lambda 2.1661\,\mu$m and $H_2$ $\lambda 2.1228\,\mu$m position-velocity curves along the major axis of the highly obscured nuclear region of NGC 253 (Figure 5a). The zero position for the velocity curve is chosen from the continuum peak. We measure a systemic velocity $V_{\rm LSR} = 235 \pm 11\,{\rm km\,s^{-1}}$, in agreement with $V_{\rm LSR} = 229\,{\rm km\,s^{-1}}$ from CO data (Canzian et al. 1988) and the optical velocity $248 \pm 10\,{\rm km\,s^{-1}}$ (Ulrich 1978). On the SW side of the nucleus, the two near-IR rotation curves are seven times steeper than the [N II] $\lambda 6584$Å optical rotation curve from Ulrich (1978) (see Figure 5a). We attribute this difference to heavy dust extinction to the SW of the nucleus (Rieke & Low 1975; Scoville et al. 1985; Walker et al. 1988; Piña et al. 1992; Sams et al. 1994). A [N II] and H$\alpha$ position-velocity curve with better spatial and velocity resolution (Muñoz-Tuñon et al. 1993) shows several dips indicative of extinction near the nucleus.

The $^{12}$CO (J = 1 $\rightarrow$ 0) 115 GHz position-velocity curve (Canzian et al. 1988) is superposed on our near-IR data in Figure 5b for the inner 40$''$ . It agrees well with the near-IR kinematics, confirming that the wavelength dependence of the kinematics is due to dust extinction. We estimate from the $0.6\,{\rm km\,s^{-1}\,pc^{-1}}$ slope in Br $\gamma$, assuming circular motions, a dynamical mass within the central 20$''$ diameter region of $4 \times 10^9\,M_\odot$ (50 times greater than would have been obtained using the [NII] lines).

The velocity dispersions at the nucleus position are 63$\pm$ 4 km s$^{-1}$ and 61$\pm$ 4 km s$^{-1}$, for Br$\gamma$ and $H_2$ respectively (corrected for instrumental broadening). This is comparable to the velocity dispersion of the gas in the nucleus of M82, determined from the Br$\gamma$ 2.16 $\mu$m emission line (Gaffney et al. 1995).

Along the minor axis of NGC 253, it has been observed that optical lines are splitted, with the velocity separation ranging from about 250 km s$^{-1}$ at position nearest the nucleus to about 450 km s$^{-1}$ further out. This splitting has been attributed to shock excitation by a starburst-driven superwind at the surface of a cone or bubble by Heckman et al. (1990) and Schulz & Wegner (1992). In the inner $\pm$10 $''$ there are no indications of velocity splitting in optical lines, in agreement with the fact that no splitting is seen in the infrared Br$\gamma$ and $H_2$ intensity profiles along the minor axis shown in Figures 3 and 4.

### 3.2. Modeling the dusty rotation curves

A quantitative numerical model is applied in order to investigate how dust could be responsible for the difference between the optical and near-IR velocity curves. The model is an expanded version ofthe Triplex model by Disney et al. (1989). It consider an inclined galaxy with an exponential distribution for the light emitting gas, embedded in an exponential distribution of dust.

Along the line of sight (z), the density of gas is given by:

$$\rho'_g(z) = \rho_g e^{-h/h_g} e^{-r/r_g} \qquad (1)$$

where $h_g$, $r_g$ are the scale height and scale lenght of the gas respectively. A similar expression is assumed for the dust distribution where $h_d$, $r_d$ are the scale height and scale lenght of the dust respectively.

At every point, it is assumed that the gas has a velocity distribution which has an average velocity given by an input rotation curve, a velocity dispersion of 10 km $s^{-1}$, and is isotropic. The galaxy is supposed to rotate cylindrically, and for the intrinsic rotation curve we have chosen a curve that rises linearly from the center to 150 km $s^{-1}$ at $4''$, and remains constant at larger radii. We choose the central optical depth in V, and convert that to the optical depth at other wavelengths using the Galactic Extinction Law (Rieke & Lebofsky 1985). For the inclination of the galaxy we take an angle of 78.5° (Pence 1981). The projected rotation curve is finally determined by taking the velocity of the peak intensity of the line of sight velocity distribution at each position along the major axis.

The scale length of the gas can be taken as a free parameter, since it is very difficult to measure from the data. Since however very few models fit the data, this method provides us a way to measure it. The same holds for its scale height. We find well-fitting solutions for a scale length of $5''$ and a scale height of $1''$. We also have to make a guess for the distribution of the dust. A conservative assumption is that it follows the distribution of the stars (e.g. Kaylafis & Bachall 1987) with a scale length of $40''$ and a scale height of $4''$ (Scoville et al. 1985).

We have plotted the projected rotation curve of the model in Figure 5a. We present rotation curves corresponding to 0, 2, 20, 30 and 40 magnitudes of extinction in the galactic center. The curve for 2 mag of extinction fits the near-IR data well, while the [N II] data are fitted well, in the SW, by 30–40 mag of central extinction in the galactic nucleus.

Hence, we estimate that approximately 35±5 mag of visual extinction in the center is required to match the SW side of the optical rotation curve, which may be compared with the value of $A_V = 24$ mag quoted by Sams et al. (1994). We therefore also attribute the agreement between the optical and near-IR rotation curves in the NE to the presence of much less local dust obscuration (Scoville et al. 1985).

### 3.3. Ionized and Molecular gas distribution

The Br $\gamma$ emission is strongly peaked at the nucleus and we detect it to $12''$ SW and $18''$ NE along the major axis (Figure 6). The $H_2$ 1–0 S(1) flux at the nucleus is three times less than the Br $\gamma$ flux there, while $\pm 8''$ away from the center the $H_2$ is stronger than the Br $\gamma$. We do not detect Br $\gamma$ further out, while the $H_2$ emission extends to $\pm 20''$ from the center along the major axis. This distribution of the ionized and molecular hydrogen gas in NGC 253 is in agreement with Circular Variable Filter (CVF) spectroscopy with a $19\farcs6$ aperture (Wright & Joseph 1989). We detect $H_2$ emission over an $18''$ range and Br $\gamma$ over an $11''$ range along the minor axis.

The CO molecular gas emission (Canzian et al. 1988) is coexistent with the ionized and molecular hydrogen emission along the major axis of NGC 253 (Figure 6). The agreement of the kinematics of the ionized and excited molecular hydrogen with that of the CO suggests that the three species coexist in the same spatial region.

The starburst galaxies NGC 253, M82, and NGC 4945 have remarkably similar far-IR and radio continuum luminosities (Rieke et al. 1980; Koornneef 1993; Ghosh et al. 1991). All three show two-lobed superwind outflow along their minor axes (Heckman et al. 1990; Schulz & Wegner 1992) and in all cases the $H_2$ emission is considered to be shock-excited by supernova remnants (SNRs) rather than by UV fluorescence (for NGC 253 see Mouri 1994; for M82 see Lester et al. 1990; for NGC 4945 see Koornneef 1993 and Moorwood & Oliva 1994).

The spatial variation of the $H_2$/Br $\gamma$ ratio however is quite different for the three galaxies (Figure 7). In NGC 253, the ratio is 0.22 in the nucleus and grows quickly with increasing radius, reaching $\approx 2$ at $10''$ radius. While the $H_2$/Br $\gamma$ ratio is only about 0.15 in the center of M82, it rises to 0.55 at the edge of the starburst core (Lester et al. 1990). In NGC 4945, the ratio remains $\approx 2.5$ in the central $10''$ and grows to $\approx 9$ further out (Moorwood & Oliva 1994). The distribution of the ionized and molecular gas in NGC 253 is qualitatively different from that in the starburst galaxies M82 (Lester et al. 1990) and NGC 4945 (Koornneef 1993; Moorwood & Oliva 1994). One explanation is that the varied behavior of the $H_2$/Br $\gamma$ ratio among the three galaxies is due the dependence of $H_2$/Br $\gamma$ on the geometry of the emitting regions (Puxley et al. 1990) as well as on the evolutionary phase of the starburst (Doyon et al. 1994). Because the ratio of H II regions and supernovae is different depending on the age and geometry of the starburst, a large range in $H_2$/Br $\gamma$ among starburst nuclei is expected (Moorwood & Oliva 1988, 1990; Puxley et al. 1990; Doyon et al. 1994).

In order to study the excitation mechanism of $H_2$ the spatial variation of the intensity ratio of $H_2$ $\lambda 2.1218\,\mu$m to $^{12}$CO ($J = 1 \rightarrow 0$) emission along the major axis of NGC 253 is shown in Figure 8. (The $H_2$ data have been convolved to the same resolution as the CO.) The ratio decreases sharply with radius, and its behavior is quite similar in M82 (CO from Lo et al. 1987; $H_2$ from Lester et al. 1990). The ratio has been normalized to unity at the nucleus in both cases. Smith et al. (1991) have shown the $H_2$ line strength depends on the shock model in molecular clouds. To this extent the integrated intensity of the $^{12}$CO emission line measures the $H_2$ column density (e.g., Scoville et al. 1987). If the $H_2$ is predominently shock-excited then we suggest that the variation in the $H_2$ $\lambda 2.1218\,\mu$m to $^{12}$CO ratio may reflect changed fraction of shocked gas due to SNR. If the same excitation mechanism for the $H_2$ can be applied to NGC 253 and M82, the $H_2$/CO ratio (Figure 8) shows that shocks may be most common near their galactic nuclei. These conclusions however are speculative and model-dependent.

## 4. CONCLUSIONS

Near-infrared observations along the major axis of NGC 253 have confirmed that the observed optical kinematics are notably affected by extinction. We attribute the steeper central gradient of the infrared velocity curve to the smaller line-of-sight optical depth, and thus to a deeper penetration into the dusty galaxy. Hence, the true mass in the rapidly rotating nucleus only becomes apparent at infrared or longer wavelengths affected minimally by dust extinction.

The spatial variations of the $H_2$ 1–0 S(1) to $^{12}$CO ($J = 1 \rightarrow 0$) ratio along the major axes of NGC 253 and M82 are similar and if the $H_2$ is considered to be principally shocked-excited then the $H_2$/CO ratio may trace the fraction of shocked gas due to supernova remnants, increasing towards their nuclei. The spatial variation of the $H_2/Br\gamma$ ratio differs among starburst nuclei, reflecting possibly differences in the geometries of the emitting regions and the ages of the starbursts.

We thank E.Perez, J.Acosta, L.Cuesta, R.Genzel and specially Alicia Gordillo for valuable comments on the manuscript. F.Prada is the holder of a SERC research studentship, at QUB, under the IAC-SERC agreement.

**Figure Captions**

**Fig. 1** Br$\gamma$ intensity profiles at different positions along the major axis of NGC 253

**Fig. 2** $H_2$ 2.12 $\mu$m intensity profiles at different positions along the major axis of NGC 253

**Fig. 3** Br$\gamma$ intensity profiles at different positions along the minor axis of NGC 253

**Fig. 4** $H_2$ 2.12 $\mu$m intensity profiles at different positions along the minor axis of NGC 253

**Fig. 5** (a) Measured radial velocities in Br$\gamma$ $\lambda 2.1661\,\mu$m and $H_2$ 1–0 S(1) $\lambda 2.1218\,\mu$m along the major axis (p.a. 52°) of NGC 253, compared with the [N II] $\lambda 6584$ Å position-velocity curve by Ulrich (1978) and the CO rotation curve by Canzian et al. (1988). Model rotation curves including the effects of dust are drawn for a central extinction of 0, 2, 20, 30 and 40 magnitudes. The steepest curve corresponds to no extinction and the shallowest to 40 mag. The portion of the near-IR curve with low radial velocity to the NE between $10''$ and $25''$ from the centre, can be attributed to a real effect, e.g. the dynamical influence of the bar (see Canzian et al. 1988) (b) The Br$\gamma$ and $H_2$ position-velocity curves are compared to the CO rotation curve from Canzian et al. (1988). The near-infrared curves show a slope of $0.6\,\mathrm{km\,s^{-1}\,pc^{-1}}$. $1'' = 16.5\,\mathrm{pc}$.

**Fig. 6** Comparison of the distribution of the Br $\gamma$ and vibrationally excited $H_2$ $\lambda 2.1218\,\mu$m emission with the $^{12}$CO ($J = 1 \rightarrow 0$) emission (Canzian et al. 1988) along the major axis of NGC 253. The CO fluxes are appropriate for an aperture of 4.″4 on the minor axis by 2.″8 on the major axis.

**Fig. 7** The spatial variation of the $H_2$/Br $\gamma$ ratio along the major axis of NGC 253 compared to that for two similar galaxies, M82 (Lester et al. 1990) and NGC 4945 (Moorwood & Oliva 1994).

**Fig. 8** The variation of the $H_2$ 1–0 S(1) to $^{12}$CO ratio is shown for the major axes of NGC 253 and M82. Multiply graphed values by 80 for M82 (based on flux in the central 12″) and by 230 for NGC 253 (based on flux in the central IR resolution element). M82 is 3.25 Mpc distant, so the scale is 15.8 pc arcsec$^{-1}$ for M82 and 16.5 pc arcsec$^{-1}$ for NGC 253.

Table 1: Observed Br$\gamma$ line flux and LSR radial velocity along the major axis of NGC 253

| Position (arcsec) | Line Flux ($10^{-14}$ erg cm$^{-2}$ s$^{-1}$) | $V_{LSR}$ (km s$^{-1}$) |
|---|---|---|
| -15.4 | 0.46±0.02 | 227.41±8.91 |
| -13.2 | 0.59±0.03 | 206.86±5.45 |
| -11.0 | 0.71±0.04 | 187.83±5.00 |
| -8.8 | 1.32±0.07 | 175.96±5.21 |
| -6.6 | 2.65±0.09 | 175.43±2.27 |
| -4.4 | 4.34±0.14 | 179.42±2.44 |
| -2.2 | 6.76±0.13 | 224.77±2.63 |
| 0.0 | 12.09±0.31 | 254.94±2.41 |
| 2.2 | 13.49±0.34 | 264.35±3.29 |
| 4.4 | 5.80±0.19 | 279.39±4.17 |
| 6.6 | 1.32±0.03 | 307.02±7.54 |
| 8.8 | 0.48±0.06 | 317.22±12.43 |
| 11.0 | 0.27±0.04 | 351.03±19.78 |

Table 2: Observed $H_2$ line flux and LSR radial velocity along the major axis of NGC 253

| Position (arcsec) | Line Flux ($10^{-14}$ erg cm$^{-2}$ s$^{-1}$) | $V_{LSR}$ (km s$^{-1}$) |
|---|---|---|
| -22.0 | 0.33±0.01 | 151.13±8.59 |
| -19.8 | 0.40±0.02 | 171.76±8.56 |
| -17.6 | 0.55±0.08 | 173.72±5.97 |
| -15.4 | 0.59±0.02 | 180.80±4.35 |
| -13.2 | 1.01±0.04 | 188.15±4.20 |
| -11.0 | 1.07±0.07 | 177.56±4.44 |
| -8.8 | 1.64±0.08 | 163.43±4.50 |
| -6.6 | 2.09±0.04 | 169.81±5.67 |
| -4.4 | 2.80±0.09 | 178.41±5.60 |
| -2.2 | 3.49±0.02 | 210.65±5.32 |
| 0.0 | 3.94±0.15 | 231.73±5.80 |
| 2.2 | 2.99±0.08 | 253.24±7.70 |
| 4.4 | 1.94±0.16 | 276.26±8.62 |
| 6.6 | 1.61±0.09 | 283.92±6.18 |
| 8.8 | 1.19±0.07 | 297.10±4.03 |
| 11.0 | 0.67±0.04 | 339.07±9.83 |
| 13.2 | 0.75±0.05 | 340.21±9.26 |
| 15.4 | 0.69±0.03 | 338.36±5.67 |
| 17.6 | 0.94±0.08 | 334.11±4.72 |
| 19.8 | 0.58±0.01 | 335.26±7.70 |

Table 3: Observed Br$\gamma$ line flux and LSR radial velocity along the minor axis of NGC 253

| Position (arcsec) | Line Flux ($10^{-14}$ erg cm$^{-2}$ s$^{-1}$) | $V_{LSR}$ (km s$^{-1}$) |
|---|---|---|
| -4.4 | 2.50±0.20 | 215.85±10.57 |
| -2.2 | 14.05±0.33 | 219.21±3.47 |
| 0.0 | 12.34±0.09 | 228.36±3.66 |
| 2.2 | 2.12±0.12 | 243.68±3.42 |

Table 4: Observed $H_2$ line flux and LSR radial velocity along the minor axis of NGC 253

| Position (arcsec) | Line Flux ($10^{-14}$ erg cm$^{-2}$ s$^{-1}$) | $V_{LSR}$ (km s$^{-1}$) |
|---|---|---|
| -2.2 | 3.46±0.18 | 217.88±6.48 |
| 0.0 | 3.79±0.08 | 194.97±6.60 |
| 2.2 | 1.99±0.08 | 191.42±4.56 |
| 4.4 | 0.83±0.07 | 185.05±8.08 |
| 6.6 | 0.58±0.06 | 195.80±14.42 |
| 8.8 | 0.36±0.06 | 225.23±14.00 |